# Pulse-echo speed-of-sound imaging using convex probes

Michael Jaeger[1,*,†], Patrick Stähli[1,*], Naiara Korta Martiartu[1], Parisa Salemi Yolgunlu[1], Thomas Frappart[2], Christophe Fraschini[2], Martin Frenz[1]

[1]Institute of Applied Physics, University of Bern, Sidlerstrasse 5, 3012 Bern, Switzerland
[2]Hologic® – Supersonic Imagine®, 135 Rue Emilien Gautier Bâtiment C, F-13290 Aix en Provence, France
[*]equally contributed
[†] Corresponding author: michael.jaeger@unibe.ch

## Abstract

Computed ultrasound tomography in echo mode (CUTE) is a new ultrasound (US)-based medical imaging modality with promise for diagnosing various types of disease based on the tissue's speed of sound (SoS). It is developed for conventional pulse-echo US using handheld probes and can thus be implemented in state-of-the-art medical US systems. One promising application is the quantification of the liver fat fraction in fatty liver disease. So far, CUTE was using linear array probes where the imaging depth is comparable to the aperture size. For liver imaging, however, convex probes are preferred since they provide a larger penetration depth and a wider view angle allowing to capture a large area of the liver. With the goal of liver imaging in mind, we adapt CUTE to convex probes, with a special focus on discussing strategies that make use of the convex geometry in order to make our implementation computationally efficient. We then demonstrate in an abdominal imaging phantom that accurate quantitative SoS using convex probes is feasible, in spite of the smaller aperture size in relation to the image area compared to linear arrays. A preliminary *in vivo* result of liver imaging confirms this outcome, but also indicates that deep quantitative imaging in the real liver can be more challenging, probably due to the increased complexity of the tissue compared to phantoms.

## 1 Introduction

Compared to other imaging modalities, pulse-echo ultrasound (US) has the advantages of being real-time, flexible, portable, non-ionizing, and low cost. Its classical grey-scale brightness (B)-mode display of tissue structure, however, often suffers from low sensitivity and specificity, limiting the diagnostic performance of US. Efforts to improve its diagnostic value have led to the development of new US-based modalities that provide complementary structural and functional disease markers in a multiparametric approach. Apart from the well-established colour flow imaging, these include elastography [1-4], and more recent techniques like photoacoustic imaging [5-7] and speed-of-sound (SoS) imaging [8-13]. In particular, the SoS is a promising disease marker, as this property can reveal changes in tissue composition and architecture that are related to variations in tissue's mass density and compressibility. In the context of fatty liver disease, for example, *in vivo* studies have shown high correlations between SoS estimations and liver fat fraction [14-16]. These studies used an autofocusing approach to determine the average SoS between the US probe and a selected point inside the liver. In order to accurately extract the SoS in liver, this technique requires a correction for the average SoS of the different tissues of the abdominal wall [14]. This can be done, for instance, by segmenting the abdominal wall and assuming its SoS.

Methods providing spatially resolved maps of tissue SoS could potentially improve liver fat fraction estimations. They inherently account for the influence of the abdominal wall [13] and could, in principle, detect liver fat fraction variations across the liver volume. In this regard, computed ultrasound tomography in echo-mode (CUTE) was introduced [17-21]. The method tracks echo positions detected under varying steering angles in transmit (Tx) [18], receive (Rx) [22], or both [20, 21]. The tracking, which can be done via phase correlation [20] or speckle tracking [23], provides spatially resolved maps of echo shifts that are related to tissue SoS. These maps allow to reconstruct tomographic SoS images using either frequency- [17, 19, 24] or space-domain [20, 21, 23] algorithms. As shown in [20], CUTE can provide reliable quantitative SoS images when a common-mid-angle (CMA) approach, with simultaneously and antidromically changing Tx and Rx angles, is combined with an appropriate forward model relating echo shift to SoS.

So far, all studies on CUTE have focused on linear probes, although convex probes are preferred for liver imaging due to their larger field of view. Compared to the linear probes, the main challenge of convex probes is the small probe aperture size in relation to the desired imaging depth. This results in a poorer angle coverage far away from the probe and may complicate the SoS reconstruction in deep regions. With the goal of liver imaging in mind, we describe an adaptation of CUTE (the new model as described in [20]) to the geometry of convex probes. The emphasis of our paper is on describing practical challenges that are specific to the curved array geometry, and strategies to make this adaptation computationally efficient. In the context of this technical paper, we also detail the calibration procedure that is an integral part of quantitative imaging but was not described before. We validate our implementation in two showcase examples of deep quantitative imaging, namely in an abdominal phantom and in a volunteer liver, and evaluate the influence of various probe geometry-specific parameters on the image quality and quantitative accuracy.

## 2 Methods and Materials

The SoS reconstruction algorithm implemented in this study builds on the methodology described for linear probes in [20]. We refer the reader to this publication for a thorough explanation of the physical and computational principles of CUTE. This section focuses on extending CUTE to convex arrays and on the required modifications to the technology.

**2. 1 Convex probe geometry and coordinate conventions**

We first introduce the conventions regarding the probe geometry and coordinate systems adopted in this work. Figure 1(a) illustrates the definition of the aperture curvature radius $R$, the angle $\alpha$ parametrising element positions on the aperture surface, the radius and azimuth $(r, \vartheta)$ defining polar coordinates of an arbitrary location in the imaging plane, and the Cartesian coordinates $(x, z)$ for lateral and axial position of the same location. We define $z = 0$ at the straight line connecting the aperture edges and $r = 0$ at the aperture surface. Even though $\alpha, r, \vartheta, x, z$ are discrete variables, we treat them as continuous throughout the manuscript for notational simplicity.

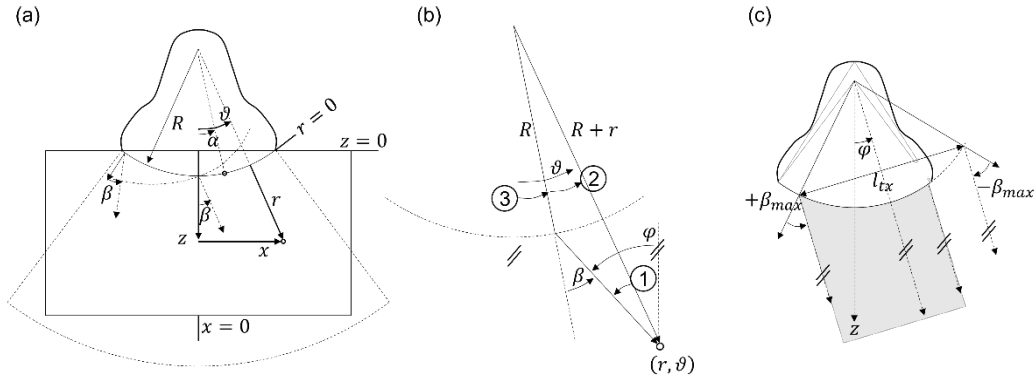

**Figure 1.** (a) Notational convention for probe geometry and spatial coordinates. The Cartesian and the polar coordinate grid are depicted as a solid rectangle and a sector area within dashed lines, respectively. A dashed curved line intersecting the aperture midpoint illustrates – for time $t = 0$ – the curved wavefront that is generated using the linear Tx delay law in Eq. 1. The wavefront is conceptually extended into the volume above the aperture surface. (b) Sketch supporting geometric derivations of Eqs. 3 and 6. i) The total Tx propagation time to a point $(r, \vartheta)$ (Eq. 3) is derived as follows: from $(\pi - \beta)$, $R$, and $R + r$, calculate the angle (1) using the sine theorem; then, use (1) and $\beta$ to calculate the angle (2); from $R$, $R + r$, and (2), calculate the propagation path length from the aperture surface to $(r, \vartheta)$ using the cosine theorem; finally, from $\vartheta$ and (2), calculate the angle (3) and set $\alpha =$ (3) in Eq. 1 to determine the Tx delay at the starting point of the propagation path. ii) Derive the Tx angle $\varphi$ relative to the $z$-axis (Eq. 6) via the alternate interior angles theorem, from angles $\vartheta$ and (1). (c) Sketch illustrating the limits of an image area (indicated in gray) that is defined for a constant Tx angle $\varphi$ relative to the $z$-axis. In this example, the image area is limited on the left side by $\beta_{max}$ and on the right side by the aperture edge.

**2.2 Data acquisition**

With linear arrays, we used plane-wave transmission for acquiring pulse-echo data. This choice ensures that – for a given Tx angle – all points in the imaging plane are insonified under similar conditions (i.e, amplitude, angle, and pulse shape) and well-defined anticipated arrival time. In comparison, e.g., a line-by-line acquisition with collimated beams has the disadvantage of the dependence of arrival time and amplitude on the position of points relative to the beam axis. With convex arrays, however, plane waves have the disadvantage that they are launched at different angles relative to the curved aperture surface, depending on the location on the aperture. Therefore, they are only actual plane waves if the tissue SoS is not only uniform (a condition also required with linear arrays) but additionally its value agrees with the *a priori* SoS used for calculating Tx delays (a condition only required for curved arrays). Else, unaccounted refraction at the probe-tissue interface leads to convergent or divergent waves.

When using a different *a priori* tissue SoS for delay-and-sum (see next section), one would thus have to take into account that the "plane" wave is no longer plane but curved. To avoid this problem, we suggest in this study to use divergent transmissions that are generated with a linear Tx delay profile, i.e., a constant delay time increment (slope) between successive elements. The delay times $t_{tx}^{(k)}$ for the $k^{th}$ transmission as a function of element position $\alpha$ are thus defined as:

$$t_{tx}^{(k)}(\alpha) = \frac{R\alpha \sin \beta_k}{c_{tx}}, \qquad (1)$$

where the index $k$ refers to individual acquisitions. Figure 1(a) illustrates the resulting curved wave front. By definition, a linear delay profile (constant slope) ensures that the propagation angle of the emitted wave relative to the aperture surface normal, measured at $r = 0$, does not depend on the location on the aperture, $\alpha$. Assuming that the SoS of the medium is uniform and has the value $c_{tx}$, this propagation angle is given by the parameter $\beta_k$. This follows from the observation that the curved aperture can be approximated as linear within a small neighbourhood around any arbitrary point $\alpha$ on the surface. Around the respective point, Eq. (1) is identical to the linear Tx delays for generating a plane wave with the angle $\beta_k$. Note that the constancy of the Tx angle relative to the aperture surface normal follows from the linear slope of the delay profile alone, and thus continues being valid even for wrong choices of $c_{tx}$ as long as the tissue's SoS is uniform. This property will become very practical further below.

In this study, we use $N_\beta = 81$ such divergent transmissions, with Tx angles $\beta_k$ ranging from -40° to 40° with a 1° angle step, .

For each $\beta_k$, we store a matrix of complex-valued radio-frequency (crf) signals $s_k(\alpha, t)$, which is obtained from the real-valued raw signals via the Hilbert transform along $t$. Although the time is discrete, and $t$ is actually an index, we keep it as a continuous dimension for notational simplicity.

**2.3 CRF image reconstruction**

For each $\beta_k$, we reconstruct a crf-mode image using delay-and-sum (DAS) beamforming, which assumes a uniform tissue SoS with the *a priori* value $c_0$. We explicitly distinguish between the *a priori* SoS for DAS, $c_0$, and the one for defining the Tx delays in Eq. (1), $c_{tx}$, because our goal is to evaluate different choices of *a priori* SoS without having to repeat the acquisition. Even though the Tx delays $t_{tx}^{(k)}(\alpha)$ cannot be changed retrospectively, it is possible to account for a $c_0 \neq c_{tx}$ in Eq. (1) by re-defining the effective $\beta_k$ that must be used for DAS. This re-definition is possible if $c_0/c_{tx}$ times the sine of the original $\beta_k$ is smaller or equal to one (else the effective $\beta_k$ does not exist). Given that – in practice – $\beta_k$ is substantially smaller than 90°, this condition holds for arbitrary choices of $c_0$ and $c_{tx}$ within the limits of realistic tissue SoS (roughly 1450 to 1650 m/s).

DAS requires a choice of the coordinate system for beamforming. In line with previous studies on linear arrays, we use in a first stepCartesian coordinates also for the present study, so that the subsequent processing steps require minimal changes compared to the implementation for linear arrays. As a second choice, however, we investigate Polar coordinates because they have several advantages over Cartesian coordinates in terms of computational and memory cost:

i) With convex arrays, the achievable lateral resolution of reconstructed echoes scales roughly proportionally to $(R + r)$. A polar coordinate grid can be matched to this resolution, so that the number of grid nodes equals the number of resolution cells. In comparison, a Cartesian grid with a sufficiently fine resolution near the aperture over-resolves the area far from the aperture and thus contains substantially more nodes than needed. DAS is therefore computationally more efficient on a polar grid than on a Cartesian one.

ii) Polar coordinates allow us to formulate the Rx propagation times from grid nodes in a translation-invariant way with respect to $\vartheta$. These propagation times therefore need to be calculated only once per $r$, either reducing computational cost when calculating them on the fly during DAS, or reducing memory cost when storing pre-calculated values.

In this study, we compare SoS image results obtained with Cartesian and polar DAS coordinates. In the following, we introduce equations in polar coordinates. For a Cartesian grid, these expressions need to be combined with a coordinate transformation from $(x, z)$ to $(r, \vartheta)$. For DAS, a crf-mode image $u_k(r, \vartheta)$ is calculated according to:

$$u_k(r, \vartheta) = \sum_\alpha s_k(\alpha, [\hat{t}_0(r, \vartheta, \alpha, \beta_k)]) \, \text{phase}(\hat{t}_0 - [\hat{t}_0]). \qquad (2)$$

Here, $\hat{t}_0$ is the anticipated round-trip time, and $[\hat{t}_0]$ is the time obtained after rounding $\hat{t}_0$ to the nearest time sample. The phase($\Delta t$) is a phase factor that compensates the timing error caused by this rounding, and is defined as phase($\Delta t$) = $exp(2\pi i f_0 \Delta t)$ with the centre frequency $f_0$. We omitted the dependency of $\hat{t}_0$ on $(r, \vartheta, \alpha, \beta_k)$ in the phase factor for brevity.

The anticipated round-trip time, which is the sum of Tx and Rx times, can be derived via standard trigonometric laws as depicted in figure 1(b). The Tx part follows:

$$\hat{t}_{0,tx}(r,\vartheta,\beta) = \frac{1}{c_0}\sqrt{R^2 + (R+r)^2 - 2R(R+r)\cos(\beta - \mathrm{asin}(R/(R+r)\cdot\sin\beta))} \quad (3)$$

$$+ \frac{1}{c_0}R\sin\beta\,(\vartheta - \beta + \mathrm{asin}(R/(R+r)\cdot\sin\beta)).$$

The first term is the propagation time from the aperture surface to $(r,\vartheta)$. As shown in figure 1(b), it takes into account the effective position on the aperture from where US waves reach $(r,\vartheta)$ under the angle $\beta_k$ relative to the aperture normal. The second term is the Tx delay according to Eq. 1 for the same effective position. Note that the origin of the time axis, $t = 0$, is defined via Eq. 1 as the time when the transmitted pulse meets the intersection of aperture surface and $z$-axis.

The Rx part is directly derived using the cosine theorem as

$$\hat{t}_{0,rx}(r,\vartheta,\alpha) = \frac{1}{c_0}\sqrt{R^2 + (R+r)^2 - 2R(R+r)\cos(\vartheta - \alpha)}. \quad (4)$$

The total round-trip time is then:

$$\hat{t}_0(r,\vartheta,\alpha,\beta) = \hat{t}_{0,tx}(r,\vartheta,\beta) + \hat{t}_{0,rx}(r,\vartheta,\alpha). \quad (5)$$

## 2.4 Coherent compounding

As described in [20], we use coherent compounding to synthetically focus/collimate the transmission with the goal to reduce clutter that stems from multiple scattering and grating lobes, and to reduce electric crosstalk. More precisely, for each angle out of a selection of Tx angles that is needed for phase tracking (see further below), we sum the crf-images over a small sub-set of $\{\beta_k\}$ that is centred at the respective Tx angle. When using a convex probe, the coherent compounding could in principle be performed along angles $\beta$ relative to the curvature normal, thus being consistent with the raw data acquisition. For a given $\beta$, however, the incidence angle relative to the radial coordinate decreases with $r$ in a non-linear way. As a result, it is not possible to choose a set of $\beta$ so that propagation angles are equidistant in any node of the image grid. Equidistant angles are, however, essential for computationally efficient implementation of the common-mid-angle (CMA) approach that is a prerequisite for reliable quantitative SoS imaging: as described in [20], using an equidistant Tx angle set allows to define CMA angles in a way that the resulting Rx angle set is identical to the Tx angle set, thus minimising the number of angles needed in the speed-of-sound inversion. To solve this problem, in this study, the coherent compounding has an additional role on top of synthetic focusing: it performs a transformation from the image set $u_k(x,z)$ or $u_k(r,\vartheta)$ (for $\beta_k$ defined relative to the curvature normal), to synthesise compounded crf (c-crf) images $u(x,z,\varphi_n)$ or $u(r,\vartheta,\varphi_n)$ for spatially uniform $\varphi_n$ (defined relative to $z$, as shown in figure 1(b)). The $\varphi_n$ can be chosen equidistantly as in [20]. As a preparation, we first write the spatially dependent propagation angle $\varphi(r,\vartheta,\beta_k)$ (derived according to figure 1(b)) as:

$$\varphi(r,\vartheta,\beta) = \vartheta + \mathrm{asin}(R/(R+r)\cdot\sin\beta). \quad (6)$$

The transformation to spatially uniform $\varphi_n$ can be expressed as

$$u(.,.,\varphi_n) = \sum_k w_c(\varphi_n - \varphi(.,.,\beta_k))\,u_k(.,.), \quad (7)$$

where $(.,.)$ stands for either $(x,z)$ or $(r,\vartheta)$. Here, $w_c$ is a symmetric weighting function that decreases with absolute difference between $\varphi_n$ and $\varphi(.,.,\beta_k)$. For each pixel, it limits the sum to those $k$ where $\varphi(.,.,\beta_k)$ is within an angular aperture around $\varphi_n$.

In this study, we choose an equidistant Tx angle set $\{\varphi_n\}$ ranging from -55° to 55° with 2.5° step size. For $w_c$ we use a Gaussian function with a $1/e^2$ radius (radius where amplitude is $1/e^2$ of the maximum) of 3°. When polar coordinates are used for DAS, a scan conversion (interpolation) from polar to Cartesian coordinates is performed after coherent compounding for each $\varphi_n$, so that following steps are performed in Cartesian coordinates independent of the initial choice of coordinates. This will be important further below for the frequency domain Rx filtering used in the phase-shift tracking step. The scan conversion was performed as follows: (i) down-conversion from polar crf-mode to IQ, by multiplying with a complex exponential of $r$ with $-f_0$ as carrier; (ii) up-sampling by a factor 2 both in $r$ and $\vartheta$; (iii) cubic interpolation onto Cartesian grid; (iv) up-conversion from IQ to crf-mode, by multiplying with a complex exponential of $r$ (defined on Cartesian grid) with $f_0$ as carrier. The down- and up-conversion was used to reduce the minimum up-sampling rate in $r$ that was required to avoid phase aliasing. It is important to point out that, in convex probes, the maximum Tx angle relative to the curvature normal, $\beta_{max}$, limits the aperture that can effectively be used for the c-crf images (see figure 1(c)). Since the aperture is curved, a 'ray' with the same $\varphi$ is emitted with different $\beta$ depending on the location on the aperture, constraining the possible emitting locations for such a 'ray'. This limits the image area that can be reconstructed for each angle $\varphi_n$, in addition to the limit imposed by the finite length of the physical aperture. For convenience, the $\beta_{max}$ can be related to a transmitting aperture length (see figure 1(c)):

$$l_{tx} = 2R \sin \beta_{max}. \tag{8}$$

The value of $l_{tx}$ represents the maximum possible diameter of the insonified area. Note that the transmitting aperture constrains the image area only where its edges lie within the physical aperture. The centre of the transmitting aperture is located where $\beta = 0$, i.e. at $\vartheta = \varphi_n$. The effective aperture results from the intersection of the physical with the transmitting aperture, and thus depends on $\varphi_n$. This differs from linear probes where the effective aperture is always identical to the physical aperture independent of $\varphi_n$. In this study ($\beta_{max}$ = 40°, R = 60.34 mm) $l_{tx}$ is 77.6 mm, whereas the physical aperture length is 61.2 mm.

## 2.5 Phase-shift tracking

At this point, the compounded crf (c-crf) mode images have been sampled on a Cartesian grid, either because the DAS was performed that way or via the coordinate transformation following the compounding. The Cartesian grid is a prerequisite for the spatial frequency domain (FD) filtering suggested in [20] (Section 3.2, p. 8 right column and p. 9 left column) for synthesising Tx-Rx-steered images from the Tx-only steered c-crf-mode images.

For a detailed description of the CMA tracking algorithm, the reader is referred to [20]. Here, we summarise the most relevant aspects for our work: i) The Rx angle set $\{\psi_m\}$ is chosen identical to $\{\varphi_n\}$ in order to minimise the computational cost of the subsequent SoS inversion. ii) We quantify the phase shift between successive pairs of antidromically changing angles $(\varphi_n, \psi_{m+1}) \rightarrow (\varphi_{n+1}, \psi_m)$. In this process, we generate Tx-Rx steered images $u'(x, z, \varphi_n, \psi_m)$ on the fly from the $u(x, z, \varphi_n)$ using FD filtering. As already in [20], tracking over antidromic sequences is in agreement with the CMA approach because the angles were chosen equidistantly. We quantify the phase shift as the argument (phase angle) of the pixel-wise Hermitian product of the successive crf-mode images. Thereby, we apply a Hann window convolution kernel in $z$ (dimension 4 mm) to the Hermitian product before calculating the argument, in order to reduce phase noise and thus improve the SNR. iii) For data reduction, the phase shift is accumulated along antidromic sequences of $\varphi_n$ and $\psi_m$, to yield phase shifts $\Delta\theta(x, z, n', m')$ between $(\varphi_{n'}, \psi_{m'+1}) \rightarrow (\varphi_{n'+1}, \psi_{m'})$, where $\varphi_{n'}$ and $\psi_{m'}$ range from -55° to 55° in 5° steps (23 angles). The finer initial angle step of 2.5° for coherent compounding has been chosen empirically with the goal to avoid phase aliasing, which occurs depending on SoS contrast when the phase shift (which is roughly proportional to angle step size) becomes larger than $\pi$. A 2.5° aperture radius is used for the directional filter.

The result of the phase tracking is a collection of 2D phase-shift maps, $\Delta\theta(x, z, n', m')$, with $n'$ and $m'$ ranging from 1 to 22. Thereby, phase-shift maps from exchanged starting and ending angle pairs, $\Delta\theta(x, z, n', m')$ and $\Delta\theta(x, z, m', n')$, would under ideal conditions be identical apart from a sign change, given the reciprocity of emitter and receiver under time-reversal symmetry. Averaging of reciprocal phase-shift maps annihilates differences between Tx and Rx beamforming, reduces noise, and it reduces the amount of data by a factor of 2, benefitting computational cost of the subsequent SoS inversion. The maps for $n' = m'$, i.e. the ones on the diagonal of the 2D collection of maps, correspond to exchanging the emitter and receiver in the transition from before to after the angle step. Therefore, $\Delta\theta$ is in theory zero, i.e., contains no information on the SoS, and these maps are not used.

## 2.6 Speed-of-sound inversion

For SoS inversion, we use the space-domain algorithm described in [20]. Note that we use slightly different notation: $c(x, z)$ is the true spatial distribution of SoS, whereas $\hat{c}(x, z)$ is the estimated one. The forward model assumes straight-ray propagation and expresses the phase shift maps $\Delta\theta(x, z, n', m')$ in terms of line integrals of the slowness deviation $\Delta\sigma(x, z)$, the difference between the inverse of $c(x, z)$ (i.e., the slowness) and the inverse of $c_0$ (the *a priori* slowness). Because the integration is a linear operation, it can be written in matrix operator notation as

$$\Delta\boldsymbol{\theta} = \mathbf{M} \cdot \Delta\boldsymbol{\sigma}, \tag{9}$$

where $\Delta\boldsymbol{\sigma}$ is a column vector containing the vectorised discrete slowness deviation, and $\Delta\boldsymbol{\theta}$ is a column vector concatenating the vectorised discrete phase-shift maps for the different combinations of $(n', m')$. As pointed out above, only those combinations are taken into account which are not equivalent or zero by time-reversal symmetry. The forward operator $\mathbf{M}$ is built from discrete line integration weights, which are defined assuming bilinear interpolation of the discrete slowness deviation along continuous straight integration lines (illustrated in figure 2(a)). For a concise description of the forward operator, the reader is referred to [20]. As in previous studies, rows in $\mathbf{M}$ corresponding to elements of $\Delta\theta(x, z, n', m')$ are set zero where the grid node $(x, z)$ is outside the image area to where ultrasound is transmitted/ from where it is received (determined by the effective aperture) for any of the angles ($\varphi_{n\prime}, \psi_{m\prime+1}, \varphi_{n\prime+1}, \psi_{m\prime}$), i.e. in areas where no echo data is available. Similarly, to mask out phase-shift noise resulting from electric crosstalk, rows in **M** are set zero that correspond to grid nodes located within the first 7 mm distance from the probe aperture (illustrated in figure 2(a)). For the convex array, a new feature is that rows corresponding to grid nodes outside the sector area are also set zero (illustrated in figure 2(a)). Moreover, columns of **M** corresponding to elements of $\Delta\sigma(x,z)$ are zero for grid nodes $(x,z)$ above the curvature radius of the aperture surface, this because the line integration weights are trivially zero in this area.

We use first-order Tikhonov regularisation [25] to well-pose the inverse problem. This regularisation constrains the first-order spatial derivatives of $\Delta\sigma$, thereby imposing spatially smooth solutions for SoS. For this purpose, we define $\mathbf{D}_x$ and $\mathbf{D}_z$ as the first-order finite difference operators along $x$ and $z$, respectively. For the convex array, the area above the curvature of the probe aperture (i.e. the area that is located outside the tissue) is excluded from this regularisation. Instead, we use zero-order Tikhonov regularisation [25] inside this area to maintain the invertibility of the forward operator. This is implemented via a diagonal matrix $\chi$ that contains ones in diagonal elements that correspond to pixels inside that area and zeros elsewhere, and by restricting the operators $\mathbf{D}_x$ and $\mathbf{D}_z$ to the pixels outside that area. Thus, the estimated slowness deviation $\Delta\hat{\sigma}$ is computed as [25, 26]

$$\Delta\hat{\boldsymbol{\sigma}} = \arg\min_{\Delta\boldsymbol{\sigma}} \|\Delta\boldsymbol{\theta} - \mathbf{M}\Delta\boldsymbol{\sigma}\|^2 + \gamma_x^2 \|\mathbf{D}_x\Delta\boldsymbol{\sigma}\|^2 + \gamma_z^2 \|\mathbf{D}_z\Delta\boldsymbol{\sigma}\|^2 + \gamma^2 \|\boldsymbol{\chi}\Delta\boldsymbol{\sigma}\|^2$$

$$= \mathbf{M}^+\Delta\boldsymbol{\theta} = \left(\mathbf{M}^T\mathbf{M} + \gamma_x^2 \mathbf{D}_x^T\mathbf{D}_x + \gamma_z^2 \mathbf{D}_z^T\mathbf{D}_z + \gamma^2 \boldsymbol{\chi}\right)^{-1} \mathbf{M}^T \cdot \Delta\boldsymbol{\theta}, \tag{10}$$

where $\gamma_x, \gamma_z$, and $\gamma$ are the regularisation parameters. Note that, because regularisation of the spatial derivative acts only on pixels outside the aperture surface, the values of $\Delta\hat{\sigma}$ in these pixels do not depend on the ones inside the aperture surface. Therefore, the choice of $\gamma$ has no visible influence on the estimated slowness deviation map, but merely enables the inversion of the bracket in Eq. 10. Finally, the map of the spatial distribution of SoS can be calculated from the slowness deviation as

$$\hat{c}(x,z) = \frac{1}{\Delta\hat{\sigma}(x,z) + \frac{1}{c_0}}. \tag{11}$$

As in previous publications [20, 21], we apply substantially stronger regularisation along $x$ than along $z$. The reason is two-fold: i) Lateral variations of SoS have a substantially stronger influence on phase shift than axial ones do, therefore lateral variations can be regularized more strongly without compromising resolution. ii) The expected anatomy in liver imaging has a layered structure. This structure is least compromised when using a stronger lateral than axial regularisation.

Given the curvature of the convex probe, it may often be reasonable to expect tissue layers of the abdominal wall and the liver surface to be bent around that curvature. In this case, the regularisation along $x$ would not optimally capture the geometry of tissue layers. As an alternative, we therefore suggest using a polar regularisation, namely of the first-order derivatives along lines of constant $r$ and constant $\vartheta$. We implement this by defining finite difference operators $\mathbf{D}_r$ and $\mathbf{D}_\vartheta$ based on $\mathbf{D}_x$ and $\mathbf{D}_z$ as

$$\mathbf{D}_r = \mathbf{D}_z \cos\vartheta + \mathbf{D}_x \sin\vartheta \quad \text{(constant } \vartheta\text{)}, \tag{12a}$$

$$\mathbf{D}_\vartheta = \mathbf{D}_x \cos\vartheta - \mathbf{D}_z \sin\vartheta \quad \text{(constant } r\text{)}, \tag{12b}$$

and by re-defining the SoS inversion analogue to Eq. (10) as

$$\Delta\hat{\boldsymbol{\sigma}} = \left(\mathbf{M}^T\mathbf{M} + \gamma_x^2 \mathbf{D}_\vartheta^T\mathbf{D}_\vartheta + \gamma_z^2 \mathbf{D}_r^T\mathbf{D}_r + \gamma^2 \boldsymbol{\chi}\right)^{-1} \mathbf{M}^T \cdot \Delta\boldsymbol{\theta}. \tag{13}$$

Note that, in this definition, $\mathbf{D}_\vartheta$ does not perform the actual derivative along the variable $\vartheta$ itself, but along the arc length of lines of constant $r$, i.e., the unit of this derivative is [m$^{-1}$] and not [rad$^{-1}$]. With this definition, the same regularisation parameters can be used for polar and Cartesian regularisation in order to provide an intrinsically fair comparison between them. This is why the same variables $\gamma_x, \gamma_z$ are used in Eq. 13 as in Eq. 10. A challenge here is that the matrices $\mathbf{D}_x$ and $\mathbf{D}_z$ must have the same size in order to be able to apply Eqs. 12. If the matrices perform a simple first order finite difference, this condition is not met, because the first order differences are defined at points half-way between the initial grid nodes in *either x or z*, respectively. In order to match the size of the matrices, the finite differences have to be defined at points centred between grid point in *x and z*. To achieve this, $\mathbf{D}_x$ performs a two-point average in $z$ together with the finite difference in $x$, and vice versa for $\mathbf{D}_z$. In this study, we compare the results for the two types of regularisation, Cartesian and polar. For the fairest comparison, we used the same definition of $\mathbf{D}_x$ and $\mathbf{D}_z$ in both approaches. It turned out that the two-point average leads to a numerical instability in the SoS reconstruction resulting in a checker-pattern SoS noise. To eliminate this pattern, the SoS images were convolved with a 2 by 2 pixel averaging kernel (2 by 2 element matrix with values 0.25).

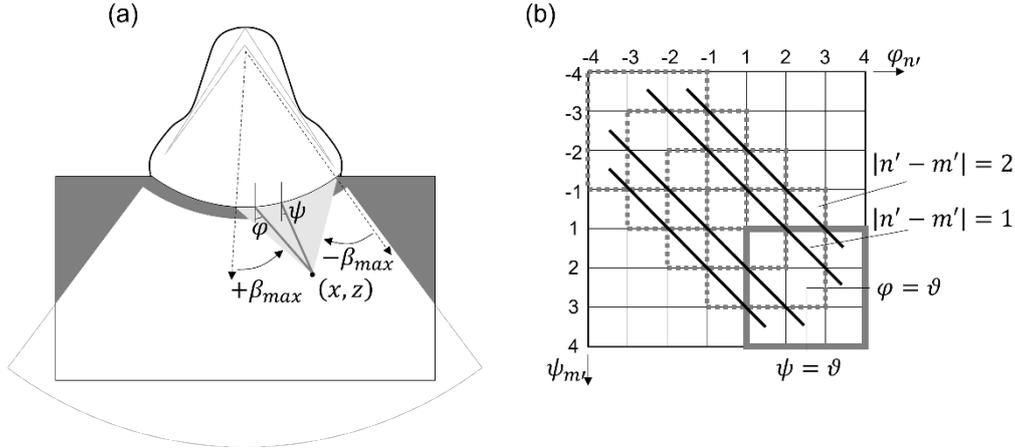

**Figure 2.** (a) Sketch of the forward model geometry. The straight integration paths for an exemplary Tx/Rx angle combination are indicated by bold dark gray lines. The area that is set zero in all phase shift maps is depicted in dark gray. This includes a 7 mm range in front of the aperture surface, and the areas outside the sector area. The angle range that can be used for a pixel at $(\vartheta, r)$, limited by $\beta_{max}$, is depicted in light gray. This is a subset of the full range for $\varphi_{n'}$ and $\psi_{m'}$. (b) Sketch illustrating the effect of the maximum angle spread $\Delta_{\varphi,\psi}$ in the 2D matrix of phase-shift maps defined by $\varphi_{n'}$ and $\psi_{m'}$. For clarity, we assume here that the full angle range is covered by 7 steps, with indices $n'$ and $m'$ running from -4 to 4, even though the actual number of steps is larger. Each small square corresponds to a phase-shift map obtained by tracking between the angle pairs corresponding to the index pairs in its lower left and upper right corner. The solid grey square indicates a subset centred around a view direction $\vartheta$. Combining corresponding sets from all view directions (indicated by dashed grey squares) results in a subset that is built from upper and lower diagonals (indicated by bold black lines). The number of diagonals is limited by $\Delta_{\varphi,\psi}$, here illustrated as the absolute difference between indices being smaller than or equal to 2.

Compared to the CUTE implementation in linear probes, we introduce again a new feature motivated by the convex probe geometry. Although the range for $\varphi_{n'}$ and $\psi_{m'}$ is substantially larger for convex than linear arrays due to the curved geometry, only part of this range can be effectively used depending on the view direction $\vartheta$ and the depth $r$, limited by $\beta_{max}$ (see figure 2(a)). We therefore introduce a parameter $\Delta_{\varphi,\psi}$ that limits the maximum allowed spread (difference) between $\varphi_{n'}$ and $\psi_{m'}$, by taking into account – in Eq. 13 – only phase-shift maps for $(\varphi_{n'}, \psi_{m'+1}) \to (\varphi_{n'+1}, \psi_{m'})$ for which $|\varphi_{n'} - \psi_{m'}| \leq \Delta_{\varphi,\psi}$. As illustrated in figure 2(b), the role of $\Delta_{\varphi,\psi}$ can be understood as follows: The sets of angles $\varphi_{n'}$ and $\psi_{m'}$ result in a 2D matrix of phase-shift maps, for all possible combinations of $n' \to n'+1$ and $m'+1 \to m'$. For a specific view angle $\vartheta$, the goal is to limit $\varphi_{n'}$ and $\psi_{m'}$ to values that are within a certain range relative to $\vartheta$, as if we were using a linear array probe pointing into direction $\vartheta$. This corresponds to choosing a square subset of the 2D matrix of phase-shift maps. Combining the subsets corresponding to different $\vartheta$ results in a subset that consists of upper and lower diagonals of the 2D matrix for which $|\varphi_{n'} - \psi_{m'}|$ is limited by a $\Delta_{\varphi,\psi}$.

We have implemented the SoS inversion for an input and output pixel resolution of 2 by 2 mm. The phase shift maps are thus downsampled to this resolution before applying the inversion.

## 2.7 Ultrasound acquisition system

We used a SUPERSONIC® Mach® 30 clinical ultrasound system for data acquisition, with a C6-1X convex probe, both provided by Hologic® – Supersonic Imagine®, Aix en Provence, France. See Table 1 for probe specific parameters. A Matlab® (R2021a, MathWorks Inc., Natick, Massachusetts, USA) based framework on a host PC generates scan sequence parameters that are sent to the system via Ethernet, and the collected data are transferred back to the host PC via the same Ethernet connection. The data processing is performed in Matlab®.

**2.8 Phantoms**

To validate the accuracy of our CUTE implementation, we use a liver imaging phantom (see figure 3) built from gelatine (Spezial Gelatine, Geistlich Pharma AG, Switzerland) , agar (A1296, Sigma Aldrich Chemie GmbH, Germany), and bakery flour (Naturaplan Weissmehl, Coop, Switzerland). It consists of four layers mimicking the human tissue anatomy. A curved indentation in the first layer matches the phantom surface to the curvature of the convex probe to allow stress-free acoustic contact. On one hand this protects the phantom from rupture, on the other hand potential stress-related variations of SoS are avoided that way. The size of the phantom is 27 cm in $x$, 15 cm in $z$, and 3 cm in $y$. The large lateral dimension ($x$) minimizes the interference of spurious reflections of grating lobes from the lateral phantom-air interface with echoes from within the image area. Similarly, the 3 cm thickness in $y$ avoids out-of-plane reflections at the phantom-air interface. Agar (2 wt% in water) was used to mimic fat layers and gelatine for the muscle layer (25 wt% in water) and the liver (20 wt% in water). Bakery flour (2 wt%) was added to both gelatine and agar to provide diffuse echogenicity. The nominal SoS values of the different layers were determined using a through-transmission time-of-flight (ToF) setup: a piston transducer with 13 mm aperture diameter, 5 MHz centre frequency (Panametrics-NDT V309, Olympus, Hamburg, Germany) is used to transmit plane pulses through a 15 mm water path and detect the echo from a plane steel reflector that is aligned parallel to the transducer aperture surface. Samples of the investigated material are prepared in flat bottom circular cuvettes (25 mm diameter, 10 mm height). After gelation, samples are removed from the cuvette and placed within the water path. After letting time for temperature equilibrium, the SoS is calculated from the ToF difference (lag of cross-corelation peak) compared to water considering the known SoS of water. This procedure yielded SoS values of 1490 m/s (fat-mimicking), 1570 m/s (muscle-mimicking), and 1555 m/s (liver-mimicking material), with an accuracy of $\pm$ 5 m/s. Published values for real tissues are around 1475 m/s for fat [27], 1575 m/s for muscle [28], and 1600 m/s down to 1500 m/s for liver depending on increasing steatosis grade [29-31].

We calibrate the system using a uniform calibration phantom to reduce the influence of differences between the model assumptions and actual experimental conditions (especially e.g., the influence of 3D sound propagation vs. 2D assumption, acoustic lense). This phantom has the same dimensions as in figure 3, but is built from gelatine (15 wt% in water) and bakery flour (2 wt%), with a nominal SoS of 1540 m/s.

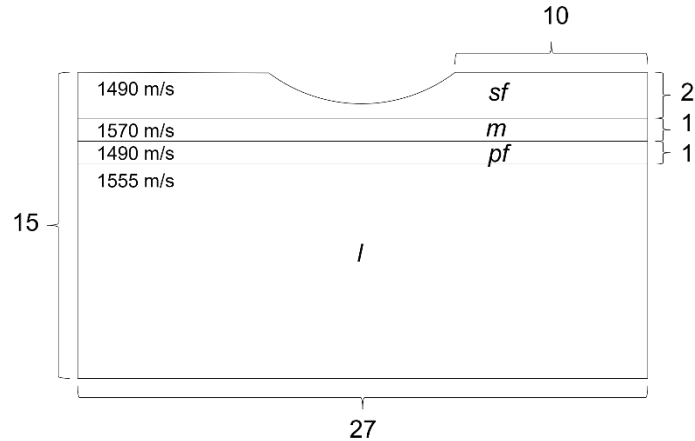

**Figure 3.** Sketch (not to scale) of liver imaging phantom, which comprises four layers mimicking the subcutaneous fat layer (*sf*), the rectus abdominis muscle (*m*), the postperitoneal fat layer (*pf*), and the liver tissue (*l*). The dimensions of the phantom and of the layers are indicated (in cm). The circular indentation in the upper phantom surface matches the probe curvature.

**2.9 Summary of parameters**

Table 1 summarises the acquisition and reconstruction parameters that are used in this study. Note that the reconstruction parameters are empirically optimised in the trade-off between computational cost and SoS image quality. A more systematic parameter optimization is beyond the scope of this study.

**Table 1**

| | |
|---|---|
| nominal centre frequency | 3.46 MHz |
| centre frequency used for acquisition | 3.0 MHz |
| probe curvature radius ($R$) | 60.34 mm |
| element pitch | 0.3360 mm |
| number of elements | 192 |
| aperture curvature length | 60.94° (64.18 mm) |
| aperture length | 61.2 mm |
| *a priori* SoS for Tx delays ($c_{tx}$) | 1540 m/s |
| *a priori* SoS for DAS ($c_0$) | 1550 m/s (phantom) 1570 m/s (volunteer) |
| Cartesian grid dimension in $x$ | 140 mm |
| Cartesian grid dimension in $z$ | 100 mm |
| crf-mode Cartesian grid resolution $x$ | 0.3360 mm (1 pitch) |
| crf-mode Cartesian grid resolution $z$ | 0.3360 mm (1 pitch) |
| polar grid dimension in $\vartheta$ | 60.94° (curvature length) |
| polar grid dimension in $r$ | 130 mm |
| crf-mode polar grid resolution in $\vartheta$ | 0.319° (1 pitch at aperture) |
| crf-mode polar grid resolution in $r$ | 0.3360 mm (1 pitch) |
| SoS Cartesian grid resolution in $x$ | 2 mm |
| SoS Cartesian grid resolution in $z$ | 2 mm |
| raw data angle range $\beta_k$ | -40° to 40° in 1° steps |
| c-crf angle range $\varphi_n$ | -55° to 55° in 2.5° steps |
| c-crf angle radius | 3° |
| phase shift kernel dimension in $z$ | 4 mm Hann window |
| phase shift angle range $\varphi_{n\prime}, \psi_{m\prime}$ | -55° to 55° in 5° steps |
| phase shift filter 1/e radius | 2.5° |
| regularisation parameter $\gamma_x^2$ in $x$ or $\vartheta$ | 20 |
| regularisation parameter $\gamma_z^2$ in $z$ or $r$ | 0.1 |
| amplitude regularisation parameter $\gamma$ | 1 |
| maximum Tx-Rx angle spread $\Delta_{\varphi,\psi}$ | 30° (6 steps of 5°) |

## 3 Results and Discussion

This section shows results obtained with the convex probe CUTE implementation using data acquired in the calibration phantom, the liver imaging phantom, and *in vivo* in a healthy volunteer. All the three examples compare implementations using polar and Cartesian coordinates for DAS and polar and Cartesian regularisation. The results are shown together with the corresponding B-mode images, which are obtained from incoherent compounding (of the squared envelope) of the c-crf images over the full range of $\varphi_n$. The SoS images are shown in their actual pixel resolution (i.e. no interpolation or other post-processing), and a consistent display range is chosen across all phantom and volunteer results.

### 3.1 Calibration phantom

Figure 4 shows results of the calibration phantom. Calibration SoS maps $\hat{c}_{cal}(x,z)$ are reconstructed using the different DAS and regularisation approaches. To reduce artefacts related to US speckle, we acquire data eight times after slightly repositioning the probe leading to different speckle realisations, and average the resulting phase shift maps. Averaging the phase shift maps – as opposed to the SoS maps – has a practical advantage: for one calibrated SoS reconstruction, only one calibration phase shift map needs to be loaded and processed instead of eight. As expected, the SoS appears approximately uniform, although we observe a slight SoS increase with depth. The absolute SoS values, however, deviate significantly from the nominal value $c_{cal}$ (1540 m/s). In our experience, this spatially dependent bias varies between probes and may result from system-specific parameters that are not accounted for in our forward model, such as the influence of the elevational profile of the acoustic lense. In all following figures, we use these calibration maps to compensate the subsequent SoS reconstructions for this bias, by applying a modified version of Eq. 11:

$$\hat{c}(x,z) = \frac{1}{\left(\Delta\hat{\sigma}(x,z)+\frac{1}{c_0}\right)-\left(\frac{1}{\hat{c}_{cal}(x,z)}-\frac{1}{c_{cal}}\right)}. \tag{14}$$

Note that, in Eq. (14), $\hat{c}_{cal}(x,z)$ is still determined according to Eq. (11). In the special case where $\Delta\hat{\sigma}(x,z)$ is measured on the calibration phantom itself, the left bracket is equal to the inverse of $\hat{c}_{cal}(x,z)$, and Eq. (14) reduces to $\hat{c}(x,z) = c_{cal}$ (i.e., uniform and quantitatively correct) as desired.

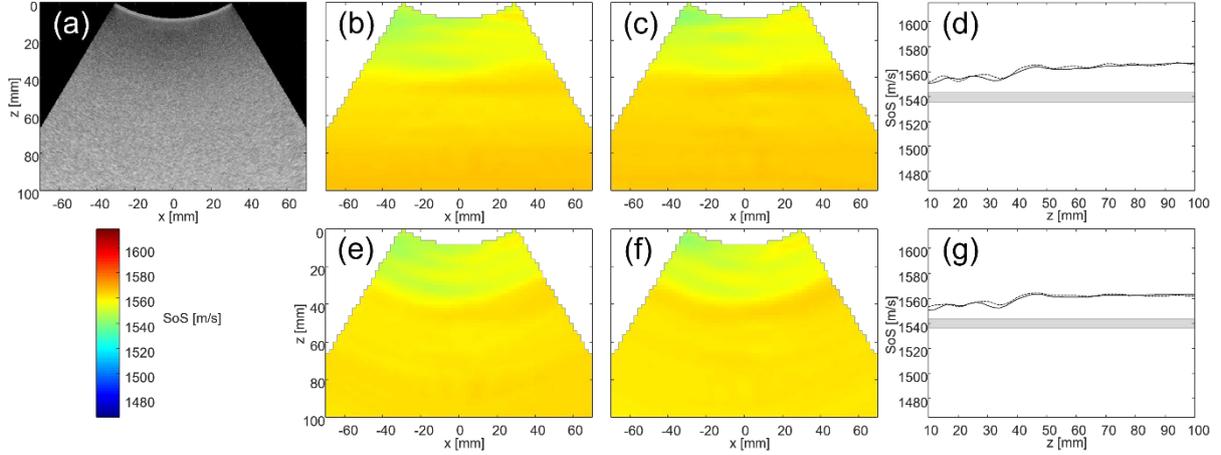

**Figure 4.** Calibration phantom results. (a) B-mode image example. (b)-(c) SoS map for Cartesian and polar DAS, respectively. Both results use *Cartesian regularisation*. (d) Corresponding depth profiles at $x = 0$ for Cartesian (solid) and polar DAS (dashed line). (e)-(f) SoS map for Cartesian and polar DAS, respectively. Both results use *polar regularisation*. (g) Corresponding depth profiles at $x = 0$ for Cartesian (solid) and polar DAS (dashed line). The bold grey lines in (d) and (g) indicate the nominal SoS value.

### 3.2 Liver imaging phantom

In this example, we show SoS reconstructions using data from the liver imaging phantom. The B-mode image shown in figure 5 allows us to distinguish the different layers of the phantom that were indicated in figure 3. When using the Cartesian regularisation, SoS images nicely reveal this layered structure (see figures 5(b) and 5(c)). Moreover, both beamforming approaches (polar and Cartesian) yield very similar results, with the polar beamforming being roughly three times faster in our implementation. The SoS inside the liver is approximately uniform, and its quantitative value is in good agreement with the nominal value. The SoS of the muscle and fat layers are slightly under- and overestimated, respectively. This may be caused by the partial volume effect (i.e., the apparent decrease of contrast due to blurring of features that are below the resolution limit) if the layer thickness is at the axial resolution limit of the SoS image.

The polar regularisation yields slightly different results (figures 5(e) and 5(f)). The different layers are well resolved, but the spatial distribution of the SoS deviates from the true layer interfaces at the lateral edges of the image domain. The interfaces are bent upwards in this area and thus disagree with the actual phantom geometry. Still, towards the centre of the domain the reconstructed layers are approximately horizontal in agreement with the actual geometry. The bending at the edges is caused by the polar regularisation favouring smoothness along lines of constant radius. Whereas the SoS value of the liver compartment is accurately reconstructed at its surface, it decreases with depth, resulting in an overall lower SoS value than with Cartesian regularisation.

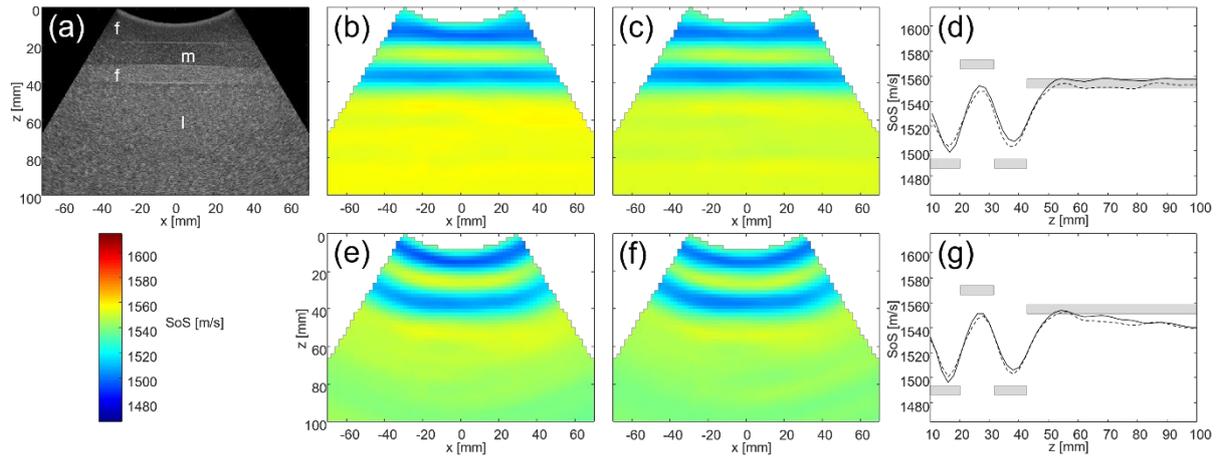

**Figure 5.** Liver imaging phantom results. (a) B-mode image (f: fat, m: muscle, l: liver). (b)-(c) SoS map for Cartesian and polar DAS, respectively. Both results use *Cartesian regularisation*. (d) Corresponding depth profiles at $x = 0$ for Cartesian (solid) and polar DAS (dashed line). (e)-(f) SoS map for Cartesian and polar DAS, respectively, both using *polar regularisation*. (g) Corresponding depth profiles at $x = 0$ for Cartesian (solid) and polar DAS (dashed line). Grey bars in (d) and (g) indicate the nominal SoS values of the phantom layers.

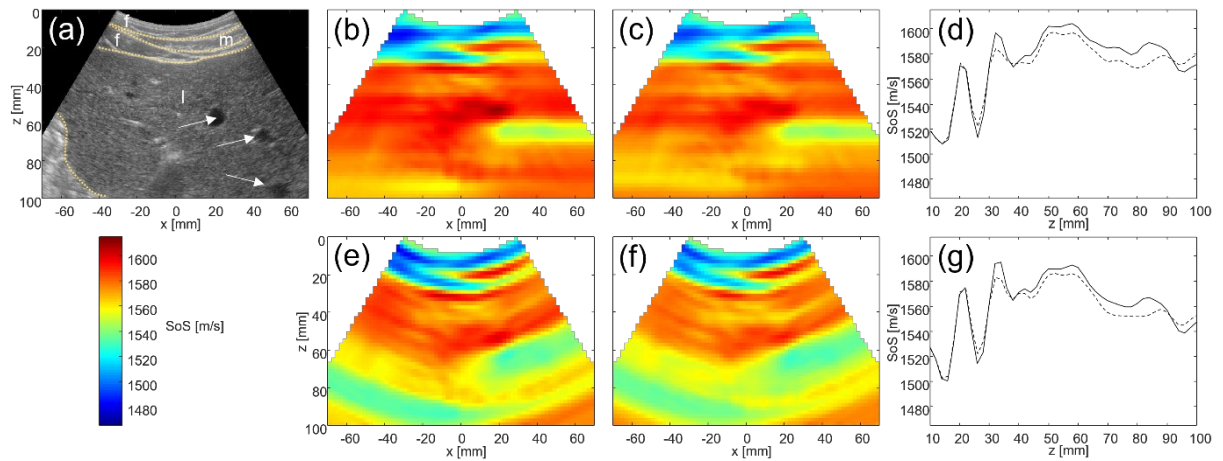

**Figure 6.** *In vivo* results in a healthy volunteer. (a) B-mode image (f: fat, m: muscle, l: liver). Dashed lines indicate tissue interfaces, arrows point at large blood vessels. (b)-(c) SoS map for Cartesian and polar DAS, respectively, with *Cartesian regularisation*. (d) Corresponding depth profiles at $x = 0$ for Cartesian (solid) and polar DAS (dashed line). (e)-(f) SoS map for Cartesian and polar DAS, respectively, using *polar regularisation*. (g) Corresponding depth profiles at $x = 0$ for Cartesian (solid) and polar DAS (dashed line).

### 3.3 In vivo results

Figure 6 shows preliminary volunteer results of imaging the liver through the ventral part of the abdominal wall. Ethics approval was obtained from Bern cantonal ethics board, ID 2020-03041. Like in the phantom example, polar and Cartesian beamforming approaches yield very similar results. The SoS images show the fat and muscle layers and capture their varying thickness along the probe aperture. Rather than as a varying layer diameter, the varying thickness is reflected as a gradually varying layer contrast, again an indication that these layers are at the resolution limit (partial volume effect). As expected from literature, SoS values in the fat are lower than in the muscle. With Cartesian regularisation, the apparent interfaces of the superficial layers do not follow the curvature of the actual interfaces observed in the B-mode image. Polar regularisation captures their appearance more accurately, because the regularisation along lines of constant radius naturally favours a curved geometry of interfaces. As already in the phantom experiment, the SoS shows the unexpected negative drift towards depth compared to Cartesian regularisation. In both approaches, the SoS in the superficial part of the liver (around 1580 m/s) agrees with reported values for healthy liver [29-31]. Unlike in the phantom results, the liver SoS becomes strongly deteriorated below roughly 50 mm depth. Here, we observe horizontal artifacts that are probably caused by the relatively large anechoic areas (blood vessels) observed in the B-mode image. Such artifacts may in the future be avoided in a more sophisticated implementation of CUTE where the absence of phase shift data within these regions is explicitly taken into account. Overall, *in vivo* results show a much higher level of SoS variations inside the liver layer compared to phantom results. This observation is in agreement with linear array results of previous studies [20, 21], and is thus not related to the convex probe implementation. The stronger variations may be caused by a higher level of reverberation clutter and stronger US aberration due to the more complex structure of real tissue compared to the phantom.

### 3.4 Role of regularisation for spatial resolution

Both phantom and *in vivo* results show artifacts in the shape of lateral or azimuthal streaks and a straightening/bending of superficial layers when using Cartesian/polar regularisation. These observations may raise doubts as to whether the layer structure of the SoS images is not simply the result of a very low lateral/azimuthal resolution, imposed by the strong lateral/azimuthal regularisation. In order to clarify this aspect, we use the same phantom as in figure 3 and acquire data with the probe tilted parallel to the imaging plane while maintaining contact between the aperture and the curved phantom surface. This is equivalent to rotating the phantom with respect to the $z$-axis, as seen in the B-mode image in figure 7(a). In figure 7(b), we show the reconstructed SoS image using polar DAS together with Cartesian regularisation. Due to the tilt, the direction of maximum regularisation ($x$) does not agree with the direction of the layer interfaces of the phantom. Still, the reconstructed SoS image follows the tilted layer geometry in the lateral middle of the image area. This demonstrates that the reconstructed layer geometry is not simply a result of the lateral regularisation. Only at the lateral edges of the image domain, the reconstructed distribution tends towards horizontal layers, resulting in a slight "s"-shape of the layer interfaces. A similar result is obtained for polar regularisation: in figure 7(c), the SoS follows the tilted plane geometry in the lateral middle of the image area, but towards the lateral edges, the reconstructed distribution is forced towards curves with constant radius. As we will discuss in the next section, the different behaviour in different areas is the result of a laterally varying data coverage: towards the lateral edges, less phase shift data is available due to reduced angle coverage than in the middle, thus the *a priori* information (horizontal layers) encoded by the regularisation term dominates the available information more in these areas.

Further, to corroborate that the observed lateral streaks are not indicative of a bad lateral resolution due to regularisation, figure 7(c) illustrates the spatially dependent point-spread functions (psf-s) related to our inverse problem. In line with a method used in geophysics [32] [26] and adapted to US tomography [33], these psf-s are defined in a theoretical way based on the linear forward model and its pseudo inverse:

$$\Delta\hat{\boldsymbol{\sigma}} = \mathbf{M}^+\mathbf{M} \cdot \Delta\boldsymbol{\sigma} \qquad (15)$$

Eq. 15 reads as follows: putting together Eqs. 10 and 14, $\Delta\hat{\boldsymbol{\sigma}}$ is the reconstructed slowness distribution resulting from the true distribution $\Delta\boldsymbol{\sigma}$. If $\Delta\boldsymbol{\sigma}$ is a point source, i.e. has the value one in a single pixel and is zero everywhere else, then $\Delta\hat{\boldsymbol{\sigma}}$ is the psf of that point source. The matrix $\Delta\widehat{\boldsymbol{\Sigma}} = \mathbf{M}^+\mathbf{M}$ thus maps point sources to psf-s, for which reason it is called the model resolution operator [26]. A point source $\Delta\boldsymbol{\sigma}$ is a unit vector containing a one at the index that corresponds to the source location. The result of multiplying $\Delta\widehat{\boldsymbol{\Sigma}}$ to $\Delta\boldsymbol{\sigma}$ is equal to the column of $\Delta\widehat{\boldsymbol{\Sigma}}$ at the column index that corresponds to the source location. In summary, $\Delta\widehat{\boldsymbol{\Sigma}}$ contains the psf-s for all possible source locations, vectorised as its columns at the corresponding indices. The fundamental assumption here is that the forward model **M** is accurate, for which reason we call the psf-s theoretical. Figure 7(d) shows a selection of these psf-s, for a grid of pixels located along 5 different radial lines with azimuth (-22°:10°:22°)at 3 different depths (15 mm, 40 mm, 70 mm). Note that, even though the appearance of layered structures is substantially influenced by the regularisation type, this influence is hardly seen in the psf-s. For this reason, we only show psf-s considering the Cartesian regularisation.

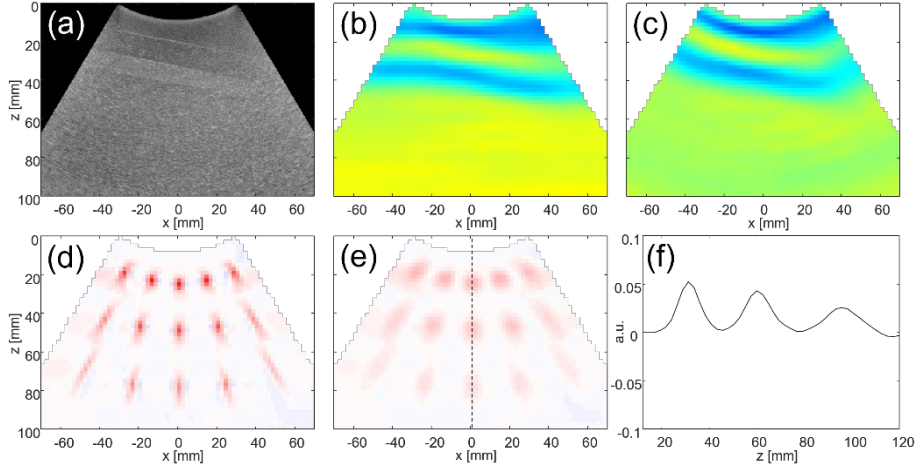

**Figure 7.** (a) B-mode image of the liver imaging phantom with tilted probe. (b) Corresponding SoS image when using Cartesian regularisation, and (c) when using polar regularisation. (cd) Spatially dependent theoretical point-spread functions related to our inverse problem. (e) Prediction of point-spread function when taking into account tracking kernel size and the Tx/Rx beam width. (f) Profile along the dashed line in (e). No colour bars are shown in these figures because the focus is on the spatial distribution profiles, not quantitative values.

The psf-s in figure 7(d) demonstrate that the azimuthal resolution of our inverse problem corresponds to roughly the pixel size independent of the position inside the image sector. This observation is reasonable: azimuthal variations (i.e. perpendicular to ray paths) of SoS are far less difficult to detect (larger phase shift magnitude) than radial variations (i.e. parallel to ray paths). The stronger lateral/azimuthal than axial/radial regularisation makes best use of this anisotropy in the available information, by regularising the SoS inversion more where it least compromises resolution. This anisotropy also explains the lack of visibility of the influence of regularisation type on the psf-s: point-like structures are easily detected due to the strong azimuthal gradient. The differences due to regularisation type are small compared to the maximum amplitude of the psf-s, but become visible via the overlap of different psf-s when 'convolving' with a layered structure.

The theoretical psf-s in figure 7(d) show the lower limit of spatial resolution determined by the inverse problem alone. In practice, the spatial resolution is additionally limited by the tracking kernel size in $z$, and by the Tx/Rx focal diameter in $\vartheta$. In figure 7(e) we approximate this by convolving the result from figure 7(d) with a 5-pixel Hann window in $z$ (mimicking the tracking kernel) and a 7-pixel Hann window (approximating the observed Tx/Rx focal profile, not shown) in $x$. Figure 7(e) suggests that the psf of our implementation is rather isotropic in an area above 50 mm depth and within ±30 mm around the $z$-axis. The lateral width of these psf-s suggest that the lateral streaks observed in the phantom and volunteer results are not the result of a low lateral/azimuthal resolution. Rather, these streaks occur due to phase shift noise. The SoS inversion generates from this noise what is least penalised by regularisation: variations that show a smaller lateral/azimuthal than axial/radial gradient.

Note that both figure 7(d) and figure 7(e) indicate a lateral variation of radial resolution: it is best near the centre of the probe aperture, and worsens towards depth and towards the lateral sector borders. Again, this is the result of the spatially varying data coverage that will be discussed in the next section. Figure 7(f) shows a profile along the $z$-axis through the psf-s in the lateral middle of the image area. The full-width half-maximum of the profile at 30 mm depth indicates an axial resolution of our implementation of roughly 10 mm in the superficial area of the SoS images, supporting the hypothesis from earlier figures that the layers in this area were at the resolution limit.

### 3. 5 Role of maximum angle spread

In Section 2.6, we have introduced a novel feature of the SoS inversion motivated by the convex array geometry, i.e., a limitation to the maximum spread $\Delta_{\varphi,\psi}$ between Tx and Rx angles $\varphi_n$, and $\psi_m$, for which we take into account phase shift data. Figure 8 illustrates the influence of $\Delta_{\varphi,\psi}$ on the angle coverage and the final SoS image. We define the angle coverage for each pixel as the total number of phase-shift maps containing data for that pixel. The spread $\Delta_{\varphi,\psi}$ is defined in numbers of the 5° angle step size that was used for the final phase-shift maps. Given that the size of the angle sets is 23, 22 by 22 phase-shift maps can in principle be obtained. However, after neglecting the 22 on the diagonal that contain zero phase-shift data (tracking between switched Tx/Rx angles) and averaging the ones that are equivalent by transmit/receive reciprocity, a maximum total of 231 independent phase-shift maps are available. Out of these, many contain no phase-shift data because the angle spread is too large given

$\beta_{max}$. Figure 8(a) reveals for each pixel the maximum possible coverage. For a maximum spread of 6 angle steps (same as in previous results) (figure 8(b)), the angle coverage is in large part of the image area equal to the maximum possible value. It only decreases slightly for superficial pixels that are closer than 25 mm to the aperture surface. Thus, the corresponding SoS images reveal hardly any difference in the axial resolution of the upper phantom layers (figure 8(e) and 8(f)). For a maximum spread of 3 angle steps (figure 8(c)), the angle coverage is substantially reduced in the area covering the upper phantom layers. As a result, the corresponding SoS image (figure 8(g)) shows a slight axial blurring and a reduced contrast. Based on this analysis, we decided that a maximum spread of 6 angle steps is a good compromise between image quality and computational cost. Note that all the angle coverage maps (figures 8(a) to 8(c)) reveal a lateral variation due to the limited probe aperture. This results in a lateral variation of the strength of regularisation relative to data availability, and explains the lateral variation of axial resolution and the straightening/bending of reconstructed layer boundaries discussed earlier.

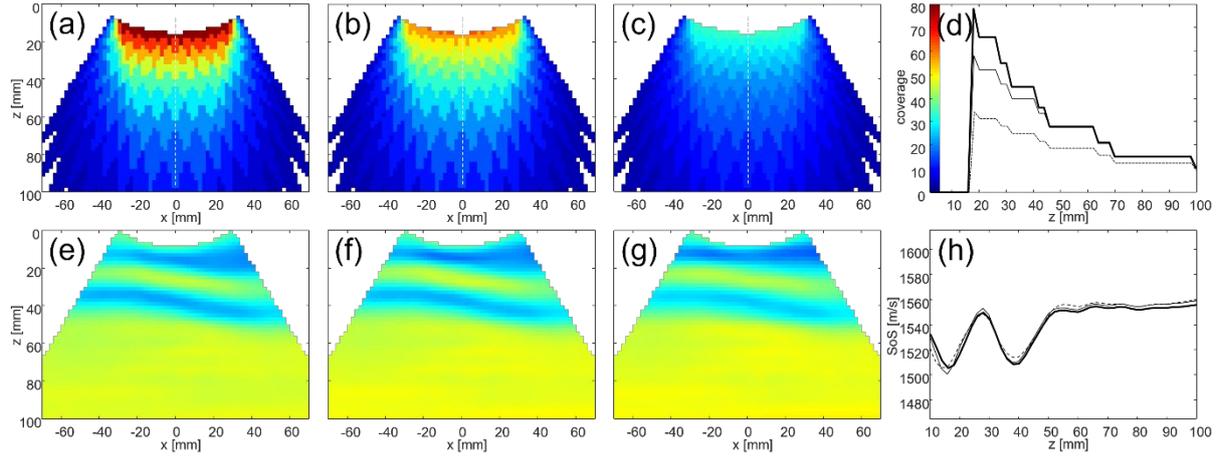

**Figure 8.** (a)-(c) Maps of angle coverage (number of phase-shift data available per pixel) for a maximum spread $\Delta_{\varphi,\psi}$ of 22, 6, and 3 angle steps, respectively. The colour bar is shown in (d). (d) Depth profiles of angle coverage for maximum spread of 22 (bold solid), 6 (solid), and 3 (dashed) angle steps. The location of these profiles is indicated with dashed white lines in (a) to (c). (e)-(g) SoS images corresponding to (a)-(c), respectively. No colour bar is shown because the focus is on the spatial distribution profile, not quantitative values. (h) Depth profiles of SoS images obtained with $\Delta_{\varphi,\psi}$ equal to 22 (bold solid), 6 (solid), and 3 (dashed) angle steps.

### 3. 6 Role of the a priori speed of sound

The first step of CUTE is to DAS the crf-signals using an *a priori* uniform SoS $c_0$. The chosen value inherently determines the detected phase shift and thus the reconstructed slowness deviation. It is accounted for in Eq. 11, which assumes that the reconstructed slowness deviation is additive to the initial guess. In addition, the calibration in Eq. 14 is expressed assuming again an additive correction to the slowness deviation. Here, this assumption of linearity/additivity is experimentally validated. Figure 9 shows SoS images obtained for the liver phantom using different values of $c_0$ ranging from 1510 m/s to 1590 m/s, using the Cartesian regularisation. For each $c_0$, the calibration was performed using calibration maps that were obtained with the respective same $c_0$. This was required because the calibrated bias depends on $c_0$. The final SoS images after calibration reveal that the influence of $c_0$ on the final SoS values is negligible over a larger range of $c_0$ values. The main visible differences are the horizontal bar artifacts inside the liver compartment. These artifacts are least pronounced when $c_0$ is in-between the SoS of the different layers and become stronger when $c_0$ deviates from that value. This effect is probably related to Tx/Rx focusing of the c-crf images: for a $c_0$ that is closer to the average SoS of the layers above the liver compartment, the Tx/Rx focusing is better inside the liver compartment, resulting in less phase-shift noise. For previous phantom results, we used $c_0$ = 1550 m/s which led to an intermediate level of artifacts.

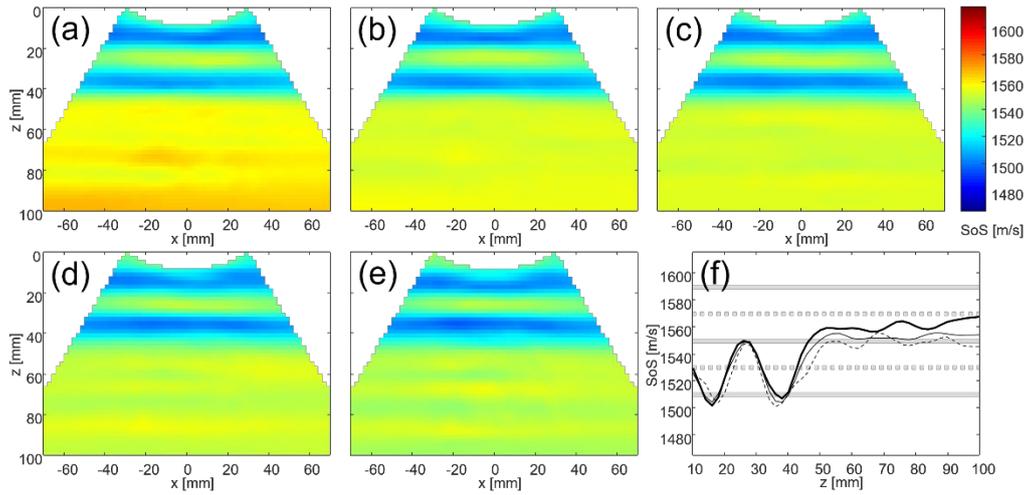

**Figure 9.** (a)-(e) Reconstructed SoS maps for liver phantom using *a priori* SoS values of 1510 m/s, 1530 m/s, 1550 m/s, 1570 m/s, and 1590 m/s, respectively. (f) Depth profiles of SoS for 1510 m/s (solid bold), 1550 m/s (solid), and 1590 m/s (dashed). The values of the *a priori* SoS are indicated by gray lines.

We also analyse the effect of $c_0$ on the *in vivo* SoS images, shown in figure 10. Contrary to what we observe in the phantom, the comparison of these images show a strong dependence of the reconstructed SoS on $c_0$. Whereas the reconstructed SoS in the superficial part of the liver (<40 mm depth) is quite robust, it varies by up to 100 m/s below 60 mm. On the one hand, this difference in robustness between superficial and deep tissue can be caused by the differences in the angle coverage between the two regions. Poorer angle coverage increases the sensitivity to phase-shift noise. On the other hand, errors due to wavefront aberrations caused by superficial layers are likely to cause more distortions in reconstructed echoes, and thus more phase-shift errors, the deeper the echoes are located. Because the real tissue is more heterogeneous than our phantom, these aberrations are stronger in *in vivo* experiments.

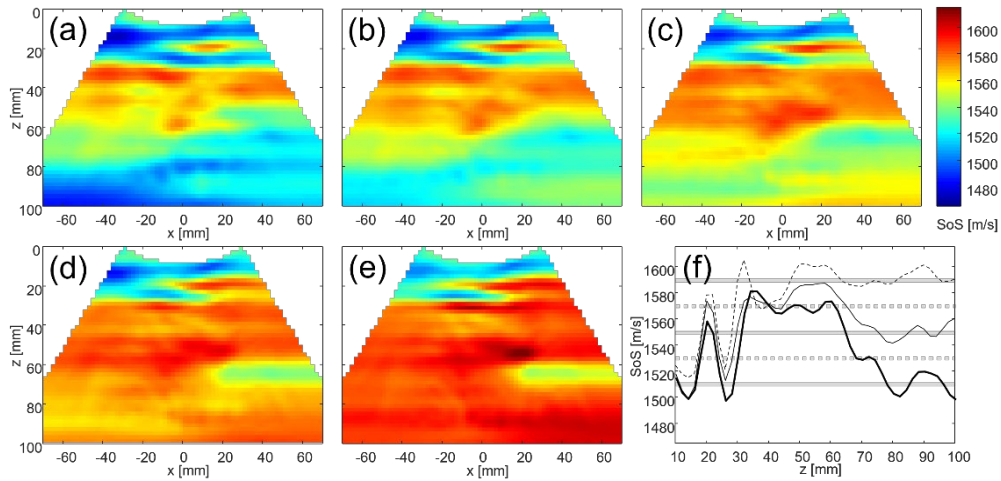

**Figure 10.** (a)-(e) Reconstructed SoS maps for *in vivo* liver data using *a priori* SoS values of 1510 m/s, 1530 m/s, 1550 m/s, 1570 m/s, and 1590 m/s, respectively. (f) Depth profiles of SoS for 1510 m/s (solid bold), 1550 m/s (solid), and 1590 m/s (dashed). The values of the *a priori* SoS are indicated by gray lines.

## 4 Conclusion

In this study, we have shown that our implementation of CUTE for convex probes is able to reconstruct the layered distribution of the SoS in the liver imaging phantom. Reconstructed absolute SoS values in the liver-mimicking part of the phantom are accurate and robust against varying the *a priori* SoS value used for DAS. This result suggests that it is in principle possible to quantitatively image the SoS at a meaningful depth for liver diagnosis, despite the small size of the probe aperture relative to the image area. Moreover, SoS images show a clear distinction between the different layers, demonstrating the spatial resolving power of our technique. Note that, while we observe that the ability to resolve 10 mm thick layers is in good agreement with the theoretically predicted psf-s, a final conclusion on the achievable spatial resolution cannot be drawn based on our results. The spatial resolution depends on a subjective choice in the trade-off between spatial and contrast resolution, determined by beamforming, tracking kernel size and regularisation. The optimum point in this trade-off can vary between subjects and average optimum parameters will have to be determined by end users depending on the diagnostic application.

The spatial resolving power of CUTE is confirmed in *in vivo* results, where subcutaneous fat and muscle layers are well distinguished near the lateral centre of the image area. While liver SoS values are reasonable within the first 20 mm from the liver capsule, quantitative imaging becomes challenging at larger depths, shown by the high level of artifacts and strong dependence of the SoS value on the *a priori* SoS. Probably, this is due to the aberrations caused by short-scale SoS variations within the different tissues of the abdominal wall, a hypothesis that needs confirmation using, for instance, numerical simulations and experiments in phantoms with higher complexity. Immediate partial solutions for reducing quantitative instability may include: i) averaging of reconstructed SoS images over multiple acquisitions at slightly different probe positions resulting in different realisations of artifacts; ii) increasing regularisation; iii) the previously proposed Bayesian approach [21] that incorporates *a priori* known tissue interfaces; iv) comparing SoS values for a fixed pre-set *a priori* SoS value for DAS. To further improve robustness, future research will focus on correcting for aberrations via more sophisticated data processing and novel data acquisition approaches. In addition, it may be beneficial to restrict the phase shift data to areas where echoes are reconstructed with high fidelity, by excluding, for example, areas containing large vessels, strong clutter, or strong degradation due to aberrations.

Phantom and *in vivo* results consistently show a difficulty in retrieving the accurate location of interfaces in the image areas towards the lateral image edges where phase shift data is scarce. We interpret this as an effect of the reduced axial resolution combined with the regularisation favouring specific geometries. As an alternative to Cartesian regularisation, we have proposed a polar regularisation strategy, to provide an *in vivo* SoS image that is more consistent with the curved subcutaneous tissue layer geometry observed in B-mode US. In the phantom study, however, we observe a negative drift of SoS towards depth compared to the ground truth when using polar regularisation. It is an open question whether this may be caused by the mismatch between the reconstructed curved and the actual flat layer geometry. While beyond the scope of the present study, this hypothesis could be tested by comparing the two approaches in phantoms with curved layer geometry. Alternatively, one may want to use a weighted combination of the two strategies, favouring polar regularisation in superficial layers but Cartesian regularisation in deep tissue.

We have demonstrated that performing DAS on a polar grid provides similar results compared to a Cartesian grid, while being computationally more efficient by taking into account the depth-dependent spatial resolution of echoes. For phase-shift tracking and SoS inversion, we have so far used Cartesian coordinates. However, similar to DAS, polar coordinates would make the coordinate grids more consistent with the spatially dependent resolution. For phase-shift tracking, we used the Fourier-domain Rx steering that requires a spatially constant Tx/Rx angle set. Even though computationally less efficient, explicit Rx steering could provide the flexibility to define a set of Tx/Rx angle pairs that is in each pixel matched to the effective angle range by which this pixel is accessed, i.e. with larger angle steps near the aperture and smaller steps far from the aperture. This could in turn again improve efficiency and potentially benefit image quality.

Even with the current limitations, this study already gives confidence that quantitative imaging of the liver will be feasible using convex probes, representing a significant step towards an improved US-based diagnosis of liver disease.

## Ethical statement

This study contains a healthy volunteer example. The research was conducted in accordance with the principles embodied in the Declaration of Helsinki and in accordance with local ethics board. The participant gave written informed consent to participate in the study.

## Acknowledgements

This work has been funded by the Swiss National Science Foundation under project no. 205320-179038. We thank the reviewers for their excellent and constructive comments.

M.J. designed the study, performed algorithm development and data analysis. and wrote the manuscript. P.S. developed the phantoms and assisted with setting up the ultrasound system and with data acquisition. N.K.M. assisted with manuscript writing and helpful discussions on inverse problem theory. PS., N.K.M., P.S.Y., and M.F. assisted with data analysis. T.F. and Ch.F. provided the ultrasound system and technical support. M.F. was the Principal Investigator for this study. All authors provided critical comment, edited the manuscript, and approved its final version. The authors declare no conflict of interests.

## Data availability

Ultrasound raw data used collected and analysed for this study are available online (DOI number to be added) or from the corresponding author upon request.